\documentclass[conference]{IEEEtran}

\IEEEoverridecommandlockouts
\usepackage{cite}
\usepackage{amsmath,amssymb,amsfonts}
\usepackage{algorithmic}
\usepackage{graphicx}
\usepackage{textcomp}
\usepackage{xcolor}
\usepackage{multirow}
\usepackage{gensymb}
\usepackage{url}

\usepackage{tikz}

\newif\ifarxiv
\arxivtrue             

\newcommand{\copyrighttext}{%
  \footnotesize \textcopyright~2025 IEEE. This paper has been accepted by the 2025 International Conference on Cyber-physical Social Intelligence (CPSI 2025). Personal use of this material is permitted. Permission from IEEE must be obtained for all other uses, in any current or future media, including reprinting/republishing this material for advertising or promotional purposes, or reuse of any copyrighted component of this work in other works.%
}

\newcommand{\copyrightnotice}{%
  \begingroup
  \newsavebox{\cpybox}
  \sbox{\cpybox}{%
    \fbox{\parbox{\dimexpr\textwidth-\fboxsep-\fboxrule\relax}{\copyrighttext}}%
  }%
  \IEEEpubid{%
    \parbox{\columnwidth}{\vspace{\dimexpr\ht\cpybox+\dp\cpybox+10pt\relax}}%
    \hspace{\columnsep}\makebox[\columnwidth]{}%
  }%
  \begin{tikzpicture}[remember picture,overlay]
    \node[anchor=south, yshift=10pt] at (current page.south)
      {\makebox[\textwidth][c]{\usebox{\cpybox}}};
  \end{tikzpicture}%
  \endgroup
}

\def\BibTeX{{\rm B\kern-.05em{\sc i\kern-.025em b}\kern-.08em
    T\kern-.1667em\lower.7ex\hbox{E}\kern-.125emX}}
\begin{document}

\title{A Configurable Simulation Framework for Safety Assessment of Vulnerable Road Users}

\author{\IEEEauthorblockN{Zhitong He}
\IEEEauthorblockA{\textit{Elmore Family School of ECE} \\
\textit{Purdue University}\\
Indianapolis, USA \\
he733@purdue.edu}
\and
\IEEEauthorblockN{Yaobin Chen}
\IEEEauthorblockA{\textit{Elmore Family School of ECE} \\
\textit{Purdue University}\\
Indianapolis, USA \\
chen62@purdue.edu}
\and
\IEEEauthorblockN{Brian King}
\IEEEauthorblockA{\textit{Elmore Family School of ECE} \\
\textit{Purdue University}\\
Indianapolis, USA \\
king360@purdue.edu}
\and
\IEEEauthorblockN{Lingxi Li$^*$}
\IEEEauthorblockA{\textit{Elmore Family School of ECE} \\
\textit{Purdue University}\\
Indianapolis, USA \\
lingxili@purdue.edu}
}

\maketitle
\copyrightnotice

\begin{abstract}
Ensuring the safety of vulnerable road users (VRUs), including pedestrians, cyclists, electric scooter riders, and motorcyclists, remains a major challenge for advanced driver assistance systems (ADAS) and connected and automated vehicles (CAV) technologies.  Real-world VRU tests are expensive and sometimes cannot capture or repeat rare and hazardous events.  In this paper, we present a lightweight, configurable simulation framework that follows European New Car Assessment Program (Euro NCAP) VRU testing protocols. A rule-based finite-state machine (FSM) is developed as a motion planner to provide vehicle automation during the VRU interaction. We also integrate ego-vehicle perception and idealized Vehicle-to-Everything (V2X) awareness to demonstrate safety margins in different scenarios. This work provides an extensible platform for rapid and repeatable VRU safety validation, paving the way for broader case-study deployment in diverse, user-defined settings, which will be essential for building a more VRU-friendly and sustainable intelligent transportation system.
\end{abstract}

\IEEEpubidadjcol

\section{Introduction}

Vulnerable road user (VRU) safety remains one of the most critical considerations in the design and deployment of advanced driver assistance systems (ADAS) and connected and automated vehicle (CAV) technologies \cite{yin2023reliable, liu2023metamining,li2002automated,nazat2024xai, li2024sora}. Traditional VRUs, primarily pedestrians and cyclists, typically navigate dedicated pathways or bike lanes \cite{wang2019vulnerable}. However, recent shifts in urban mobility, particularly the rapid adoption of shared micro-mobility solutions such as electric scooters (e-scooters), have introduced new safety challenges. E-scooter riders, lacking protective structures and exhibiting unique mobility patterns, represent an increasingly significant safety risk to the modern transportation system \cite{CPSC}. The latest Euro NCAP protocol also includes powered two-wheelers, specifically motorcyclists, in its collision avoidance and safety assessments \cite{euroncap}, underscoring the need to expand traditional VRU definitions and corresponding safety-evaluation frameworks.

\begin{figure}[t]
\centerline{\includegraphics[width=0.8\linewidth]{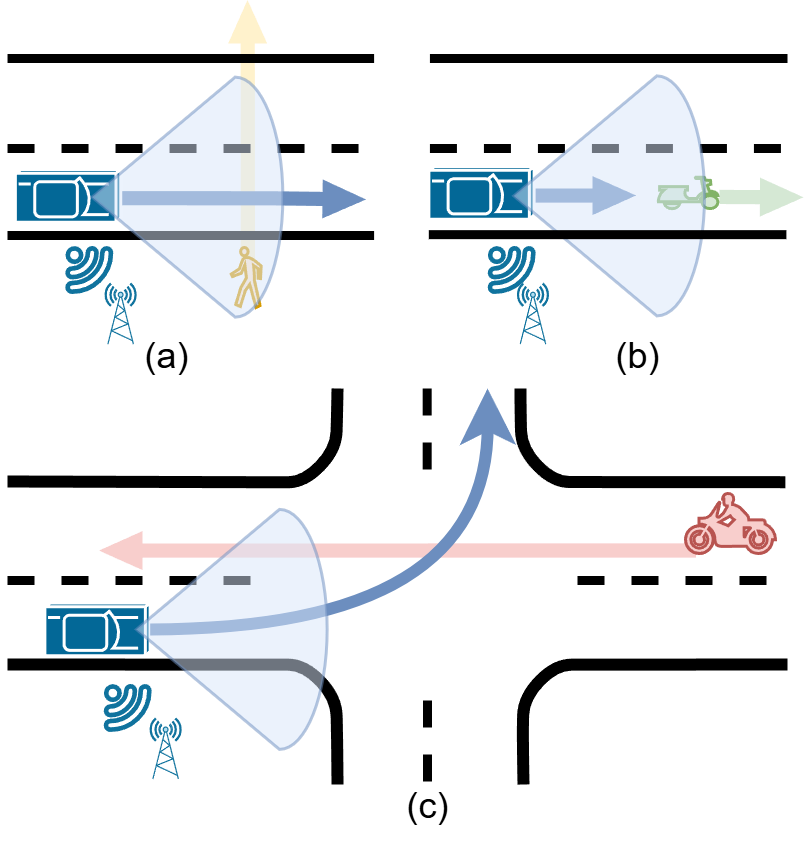}}
\caption{Three VRU testing scenarios based on the Euro NCAP VRU testing protocol \cite{euroncap}: (a) Car-to-Pedestrian Nearside Adult, (b) Car-to-Bicyclist Longitudinal Adult, (c) Car-to-Motorcyclist Front Turn across Path. Apart from the testing benchmark, a roadside infrastructure can provide additional critical VRU information by sending vehicle VRU awareness messages. Each category of VRU has different moving behaviors, e.g., varying moving directions and speeds.}
\label{fig:config}
\end{figure}

Standardized testing scenarios, such as those defined by Euro NCAP (shown in Fig.~\ref{fig:config}), offer essential benchmarks to evaluate vehicle safety systematically. These scenarios replicate realistic and challenging situations, allowing vehicle manufacturers and researchers to quantify safety performance consistently. Simulation-based frameworks have become indispensable tools, compared to costly and time-consuming real-world tests, facilitating extensive scenario testing, repeatability, and controlled evaluations\cite{xu2022safebench}.

Despite considerable advances in vehicle-VRU interaction modeling and simulation platforms, significant research gaps persist. Existing simulation environments frequently lack targeted support for dynamic VRU interactions, particularly those involving unpredictable maneuvers and sensing-limited scenarios commonly encountered with pedestrians, e-scooters, and motorcyclists. Additionally, many current frameworks provide limited flexibility in integrating and testing diverse motion-planning methodologies. These limitations hinder thorough assessments and validations of ADAS performance under realistic VRU interaction conditions.

To address these challenges, this paper aims to develop a simulation platform that can digitize standardized VRU testing scenarios within a controlled virtual environment. The framework enables the rapid prototyping of collision-avoidance logic and facilitates cyber-physical system studies by exposing related parameters during scenario generation.  Three qualitative case studies: (i) a pedestrian crossing on a straight road, (ii) a slow e-scooter leading on a straight road, and (iii) a motorcyclist crossing at an intersection, demonstrate the utility and highlight the potential of the software-defined simulation stack.

The key contributions of this paper are summarized as follows:

\begin{enumerate}
    \item Developed a simulation framework that digitizes Euro NCAP VRU tests and supports configurable perception, V2X, and motion-planning modules.
    \item Conducted qualitative validation through three classic VRU interaction scenarios, covering pedestrians, e-scooters, and motorcyclists.
    \item Implemented a rule-based collision-avoidance controller, providing a baseline for future ADAS research on VRU safety.
\end{enumerate}

\vspace{-2mm}
\section{Related Work}
\subsection{Vehicle and VRU Interaction}
\vspace{-1mm}
As automated driving systems continue to mature, one of the most complex and safety-critical domains of their operation is the interaction with VRUs, including pedestrians, cyclists, motorcyclists, and increasingly, e-scooter riders. Research in this area highlights the nuanced challenges posed by the unpredictability, variability, and contextual sensitivity of VRU behavior. The ultimate goal is to ensure that automated systems can not only detect and predict VRU movements but also engage with them in ways that are safe, understandable, and trustworthy.

Karpinski et al. \cite{karpinski2023comparison} conducted a large‑scale analysis of U.S. crash data, revealing that e‑scooter riders, while demographically similar to motorcyclists, exhibit collision patterns closer to pedestrians, particularly at unsignalized intersections and during nighttime.  Their statistical findings motivate VRU-specific control strategies in the design of automated vehicles (AVs). In \cite{saleh2017vru}, Saleh et al. introduced a probabilistic trust model in which explicit AV signals (e.g. brake light modulation) are tuned to maximize VRU confidence, demonstrating in human‑subject studies that improved transparency reduces risky crossing decisions. A related study by Chen et al. \cite{chenVRU} presented a hybrid VRU trajectory prediction pipeline that fuses a convolutional LSTM network with a microscopic Social Force Model (SFM).  The DL component captures scene context (lane geometry, traffic signals), while the SFM encodes inter-agent repulsion and goal‑directed forces.  In closed-loop tests, this hybrid method reduced endpoint prediction error by \(11\%\) relative to pure data‑driven baselines.

\subsection{Simulation Platforms and Testing Frameworks}
\vspace{-1mm}
Because real-world trials cannot safely or reliably reproduce rare or hazardous VRU encounters, simulation frameworks are indispensable. Contemporary testbeds span pure software-in-the-loop (SIL) environments, hardware-in-the-loop (HIL) rigs, and full vehicle-in-the-loop (VIL) systems.

Son et al. \cite{son2019simulation} presented a co‑simulation environment coupling Simcenter Amesim (vehicle dynamics) with Prescan (sensor and traffic scenario modeling).  Their closed‑loop setup runs at real‑time rates and supports automated scenario sweeps, enabling quantitative studies of control strategies under varying weather, lighting, and traffic densities. In \cite{marko2019scenario}, Marko et al. built on the Open Simulation Interface (OSI) standard: by defining a message schema for VRU states, they achieved plug‑and‑play interoperability among perception modules, traffic generators, and control algorithms.  The proposed modular approach reduces integration overhead when swapping sensor models or introducing new VRU behaviors. In addition to purely SIL testing methodologies, Solmaz et al. \cite{solmaz2019novel} and Shao et al. \cite{shao2019evaluating} advanced HIL and VIL testbeds, bridging the gap between virtual simulations and real-world experiments, thus ensuring realistic and reliable evaluations of vehicle control systems.

Despite these advances, existing frameworks often lack: (i) flexible scenario digitization for emerging VRU types, such as e‑scooters and motorcyclists; (ii) seamless integration of novel traffic safety technology or infrastructure, and (iii) combined evaluation metrics balancing safety, efficiency, and comfort.  Addressing these gaps is the primary focus of this paper.

\vspace{-2mm}
\section{Simulation Framework and Scenario Digitization}
\label{Sec:Framework}

\begin{figure}[t]
\centerline{\includegraphics[width=0.65\linewidth]{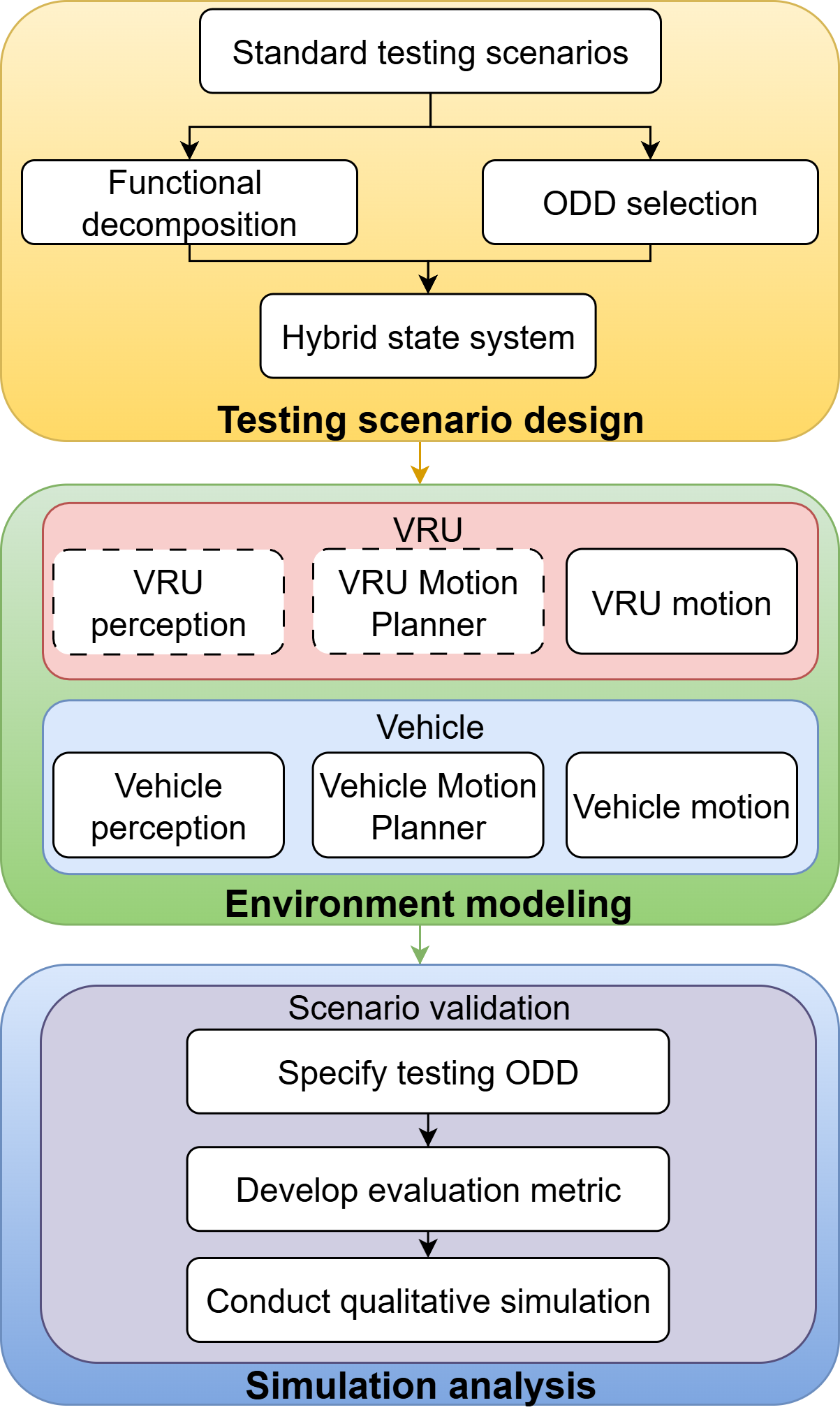}}
\caption{Proposed VRU analysis framework based on standard testing protocol and model-based traffic simulation.}
\label{fig:frame}
\end{figure}

Adapted from our prior e-scooter study \cite{he2023simulation}, the proposed framework proceeds in three stages: scenario design, environment modeling, and simulation analysis, as illustrated in Fig.~\ref{fig:frame}. 

\vspace{-1mm}
\subsection{Testing Scenario Design}
\vspace{-1mm}
The scenario design comprises two stages. First, we develop the traffic scene layout and perform a functional decomposition of key elements, such as vehicle control strategies and advanced technological features, to analyze their theoretical impact on the Operational Design Domain (ODD). Once the underlying traffic environment is instantiated, we integrate the target agents (vehicles and VRUs) into the scenario for systematic testing.
\vspace{-1mm}
\subsection{Environment Modeling}
\vspace{-1mm}
We represent both the ego vehicle and VRUs via modular components. The VRU and vehicle motion are the basic layer that drives the road users towards their destinations. Additionally, perception and motion-planning modules facilitate interactions between agents. In this paper, we consider VRUs as dummy robots used in real-life standard VRU testing \cite{euroncap} for simplicity and to illustrate proof of concept. The vehicle's sensing and motion control architecture is detailed in Section~\ref{Sec:veh_ctrl_v2x}.
\vspace{-1mm}
\subsection{Simulation Analysis}
\vspace{-1mm}
In the final stage, we apply a tailored sustainability metric, encompassing safety, comfort, and efficiency, to quantify controller performance as developed in another prior study \cite{he2024sustain}. Statistical evaluation across forward simulations validates the effectiveness of control strategies in VRU-involved scenarios. These proof-of-concept results offer actionable insights for refining ADAS/AD algorithms in subsequent development phases.

\vspace{-2mm}

\section{Traffic Simulation Models}
In this section, we specify the agent models used in all scenarios. VRUs are represented by kinematic point masses, while the ego vehicle combines a kinematic bicycle model, a rule-based finite-state planner, and an
optional V2X awareness channel.
\vspace{-1mm}
\subsection{VRU Model}
We utilized a point-mass model to represent the VRUs in the simulation environment for computational simplicity. The VRU movement can be formulated as: 
\begin{equation}
    \ddot{X}_{\rm vru}=\frac{1}{m_{\rm vru}} f_{\rm total},\label{eq:pointmass}
\end{equation} 

where
\[
\mathbf X_{\rm vru}
=\begin{bmatrix}
s_{{\rm vru},x}\\[3pt]
s_{{\rm vru},y}\\[3pt]
v_{{\rm vru},x}\\[3pt]
v_{{\rm vru},y}
\end{bmatrix}
\in\mathbb{R}^4
\quad\text{and}\quad
\mathbf f_{\rm total}
=\begin{bmatrix}F_x\\[3pt]F_y\end{bmatrix}.
\]

\(\mathbf X_{\rm vru}\) is the VRU state vector containing the position and velocity information along the longitudinal and lateral axes. \(\mathbf f_{\rm total}\) is the net force applied to the VRU.  In our implementation, we set the VRU as a dummy to move from still to the maximum speed \(v_{\rm vru, max}\) along a prescribed path according to the VRU testing protocol \cite{euroncap}.

The three VRU categories differ primarily in their top speeds and physical footprints.  Pedestrians travel slowest (around \(1.8\,\mathrm{m/s}\)), e-scooters reach up to \(13.4\,\mathrm{m/s}\), and motorcyclists up to \(30\,\mathrm{m/s}\).  We model each VRU’s occupied area as a circle: pedestrians with radius \(0.5\,\mathrm{m}\), e-scooters \(1.0\,\mathrm{m}\), and motorcycles \(1.5\,\mathrm{m}\). These dimensions also align with the real-world proportions and the standard VRU testing protocols.

\subsection{Vehicle Control and VRU Awareness Integration}
\label{Sec:veh_ctrl_v2x}
\begin{figure}[htp]
\centerline{\includegraphics[scale=0.55]{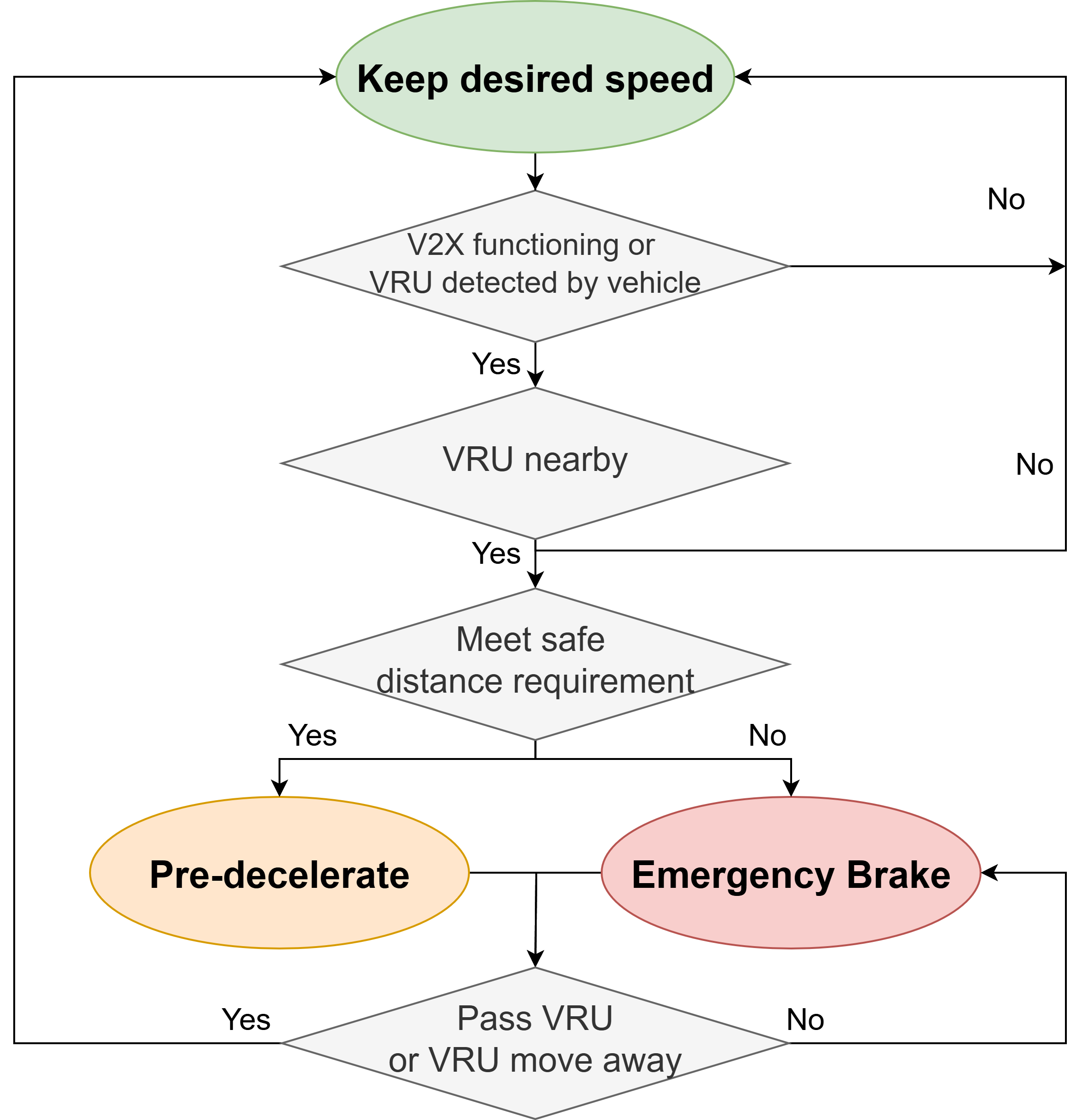}}
\caption{State transition of the collision avoidance planner integrated with the perception module.}
\label{fig:cap}
\end{figure}

The ego vehicle is a kinematic bicycle with longitudinal motion governed by the
finite-state planner shown in Fig.~\ref{fig:cap} and the steering motion follows the prescribed path.

\subsubsection{Rule-based Collision Avoidance Planner}

The FSM planner maintains a simple “cruise” behavior until a VRU is detected, either via the vehicle's onboard sensor or a VRU awareness message provided by the V2X module. The planner then evaluates the proximity of the approaching VRU and issues a gentle pre-deceleration for preventive driving purposes if the VRU is beyond the safe distance. The safe distance requirement is defined based on a two-second rule: $D_{\rm safe} = 2 \cdot V_{\rm veh}$\cite{2secrule}. Otherwise, the vehicle will implement an emergency braking command to avoid the potential collision. Once the VRU has passed, the motion planner returns to the desired cruise mode.

By structuring these modes and transitions, the FSM planner separates nominal driving, early warning, and collision‐avoidance maneuvers.

\subsubsection{V2X Module Integration}
\label{Sec:v2x}
To extend the awareness of the vehicle beyond its onboard sensors, which is defined as a sector-shaped field of view \cite{he2023vpi}, we fuse V2X broadcasts of VRU state with the perception system \cite{lobo2024adaptive}. We borrow the concept of V2X from \cite{he2024rampcast}, where the road infrastructure can provide critical traffic information to the vehicle. In our testing scenario, we assume the V2X communication is zero-latency and the connectivity is reliable. The vehicle takes actions based on the high-level control strategy by comparing the distance to the roadside unit and the VRU awareness communication range. If the V2X is disabled, the vehicle can only depend on the onboard sensor, which is represented as a blue sector. Only when the VRU is inside the sector region can the vehicle perceive the existence of the VRU. 
\vspace{-1mm}
\section{Simulation Experiments}
As shown in the VRU testing configuration in Fig.~\ref{fig:config}, the simulated traffic scenes include a straight road and an intersection. The ego vehicle may encounter a crossing pedestrian or a moving e-scooter rider with a lower velocity on the straight road. In the intersection layout, the vehicle may interact with an approaching motorcyclist. 

Two collision avoidance methods: baseline CAP, and V2X-assisted CAP will be examined using the evaluation metric proposed in the previous work. Details of the assessment can be found in \cite{he2024sustain}. In general, the safety index is compared with the safety margin based on the time-to-collision calculation. Efficiency represents the time to complete a trip, with a higher efficiency value indicating less time loss. The comfort indicator depends on the acceleration values and the derivative of acceleration, the jerk values. A smoother vehicle control strategy would lead to a greater comfort index.

\subsection{Qualitative Simulation Configuration}
\vspace{-1mm}

\begin{table}[b]
\caption{Simulation Configurations for Qualitative Analysis}
\centering
\resizebox{\columnwidth}{!}{%
\begin{tabular}{|c|ccc|c|}
\hline
\multirow{2}{*}{Symbol} &
\multicolumn{3}{c|}{Scenario Value} &
\multirow{2}{*}{Units} \\ \cline{2-4}
 & Pedestrian Crossing & E-scooter Leading & Motorcyclist Crossing & \\ \hline
$(x_{\rm veh,init},\,y_{\rm veh,init})$ & $(-50.0,\,-1.75)$ & $(-50.0,\,-1.75)$ & $(-30.0,\,-1.75)$ & m \\
$v_{\rm veh}$                            & $15.0$          & $15.0$           & $6.0$ & m/s \\
$R_{\rm veh,FOV}$                        & $20.0$          & $20.0$           & $20.0$ & m \\
$\alpha_{\rm veh,FOV}$                   & $120$           & $120$            & $120$ & \degree \\ \hline
$(x_{\rm vru,init},\,y_{\rm vru,init})$ & $(18.0,\,-10.0)$ & $(0.0,\,-1.75)$ & $(30.0,\,1.75)$ & m \\
$v_{\rm vru}$                            & $1.8$          & $13.4$           & $18.0$ & m/s \\ \hline
\end{tabular}}
\label{tab:config_qual_sim}
\end{table}

\begin{figure*}[t]
\centerline{\includegraphics[width=1.0\linewidth]{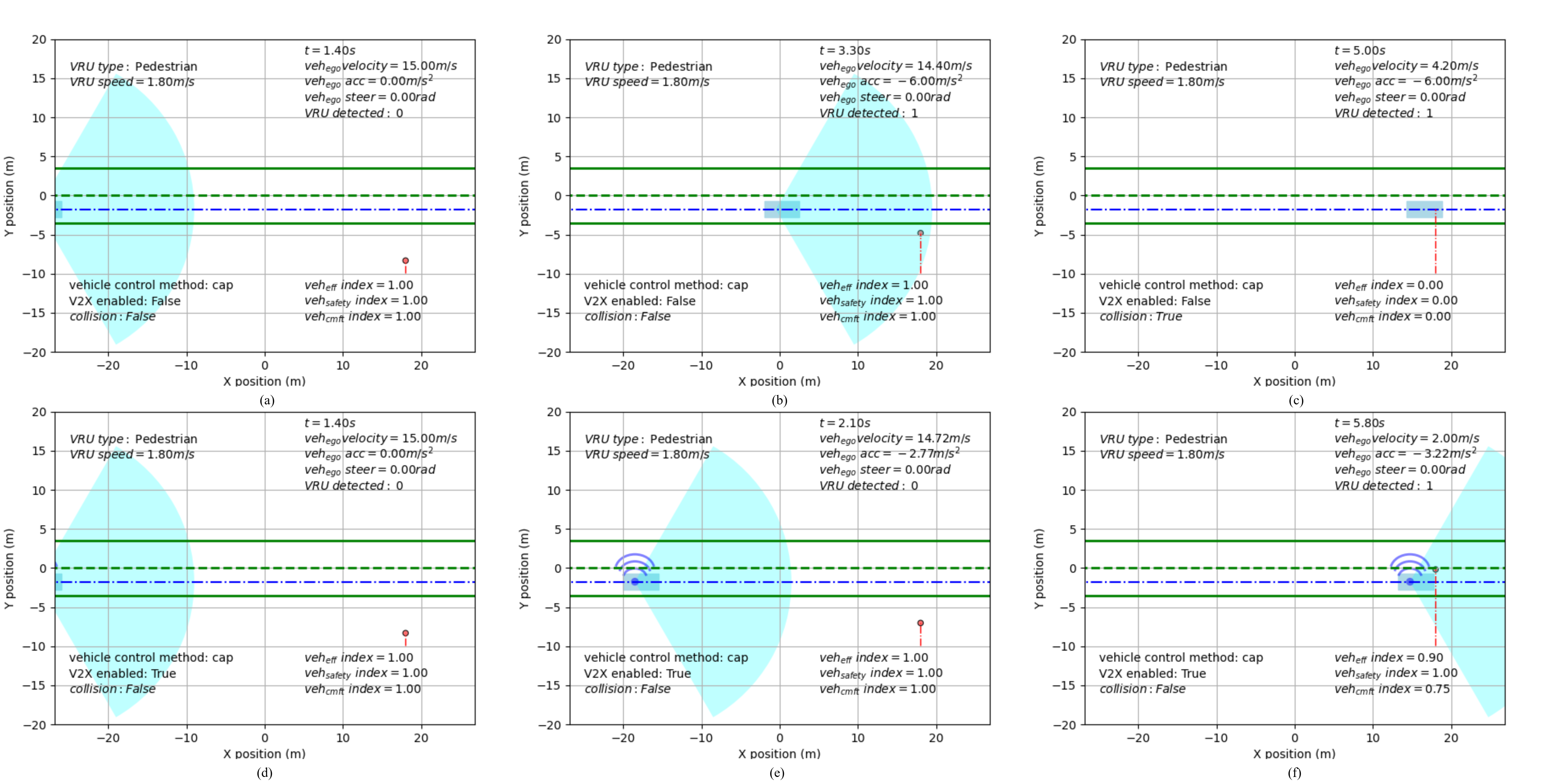}}
\caption{Snapshots of the \textbf{orthogonal pedestrian-crossing} scenario.
Top row (a)-(c): baseline CAP without V2X.  
(a) When $t=1.4\,\mathrm{s}$ pedestrian is still outside the ego vehicle’s FOV;  
(b) $t=3.3\,\mathrm{s}$: pedestrian enters the FOV and the ego vehicle begins
braking, but the remaining gap is small;  
(c) At $t=5.0\,\mathrm{s}$ the vehicle collides with the pedestrian.  
Bottom row (d)-(f): CAP augmented with V2X.  
(d) Ego vehicle approaches the pedestrian with the same initial velocity at $t=1.4\,\mathrm{s}$;  
(e) Proactive deceleration starts before detection by the onboard sensor at
$t=2.1\,\mathrm{s}$ when the ego vehicle enters the VRU awareness reception zone;  
(f) $t=5.8\,\mathrm{s}$: pedestrian clears the path and the collision is
avoided.}

\label{fig:qual_results_ped}
\end{figure*}

\begin{figure*}[h!]
\centerline{\includegraphics[width=1.0\linewidth]{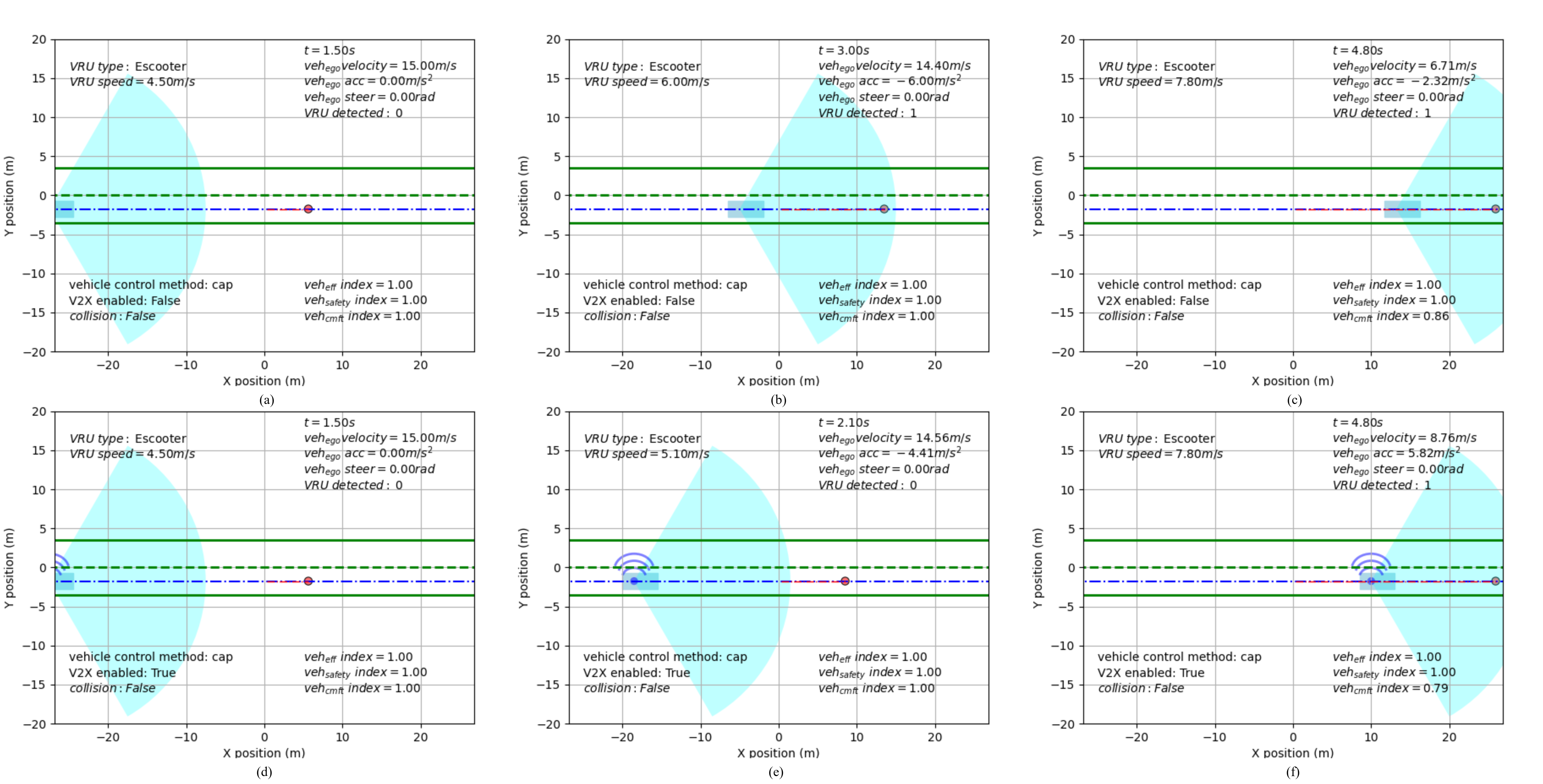}}
\caption{Snapshots of the \textbf{same-lane e-scooter following} scenario.
Top row (a)-(c): CAP without V2X. The ego vehicle closes on a slower e-scooter,
detects it at $t=3.0\,\mathrm{s}$, and reduces speed; by
$t=4.8\,\mathrm{s}$ a safe headway is established.  
Bottom row (d)-(f) demonstrates the case when V2X is enabled to assist CAP. Early warning triggers a less
gentle but more conservative deceleration profile. Both CAPs avoid collision successfully.}

\label{fig:qual_results_e_scooter}
\end{figure*}

\begin{figure*}[t]
\centerline{\includegraphics[width=1.0\linewidth]{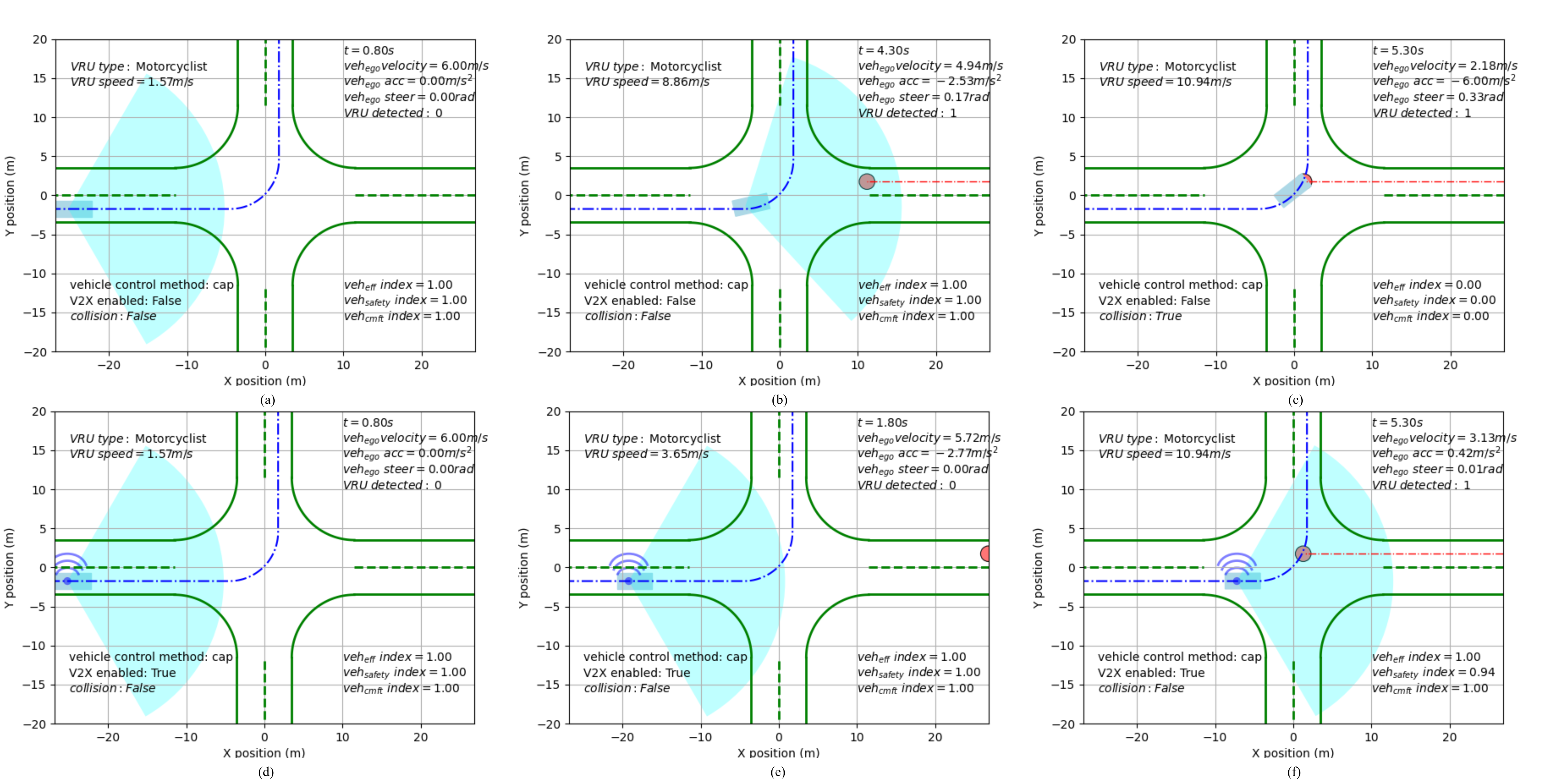}}
\caption{Snapshots of the \textbf{left-turn motorcyclist crossing}
scenario.  
Top row (a)-(c): CAP without V2X. The ego vehicle initiates the turn, detects
the oncoming motorcyclist late at $t=4.3\,\mathrm{s}$, and an impact
occurs at $t=5.3\,\mathrm{s}$.  
Bottom row (d)-(f): With the V2X-assisted CAP, the VRU message is broadcast at 
$t=1.8\,\mathrm{s}$ prompts early braking; the motorcyclist passes safely
and no collision is recorded.}

\label{fig:qual_results_biker}
\end{figure*}

The configurations of the traffic simulations for the three target scenarios are presented in Table~\ref{tab:config_qual_sim}. The vehicle perception module had the same default sensor with a \(20\,\mathrm{m}\) detection range and a \(120\,\degree\) detection angle. In the two straight road scenarios, the desired cruise speed was set as \(15\,\mathrm{m/s}\), while the moving speed was reduced to \(6\,\mathrm{m/s}\) for a more realistic turning scenario. The pedestrian in the straight road scenario would start from one side of the road and cross the road when the host vehicle is approaching, while the e-scooter would move forward in the same direction as the vehicle. The motorcyclist would cross the intersection from the opposite direction with the highest VRU speed. Controlled experiments were conducted to compare the two vehicle motion planning methods across the three scenarios.

\subsection{Qualitative Analysis and Discussion}
\vspace{-1mm}
Figures \ref{fig:qual_results_ped}-\ref{fig:qual_results_biker} collectively illustrate how early VRU awareness influenced the rule-based finite-state planner. Without V2X, the controller relied solely on its onboard sensor. In VRU fast-approaching, orthogonal conflicts (pedestrian and motorcyclist cases), vehicular detection occurred too late, and the vehicle could not reduce adequate speed before the final impact. Zero-latency V2X shifts the
\emph{Cruise} $\rightarrow$ \emph{Pre-Slow} and
\emph{Pre-Slow} $\rightarrow$ \emph{Emergency-Brake} transitions forward
by roughly one second, providing enough distance to stop or yield, and thereby eliminating collisions in both cases.

In Fig.~\ref{fig:qual_results_e_scooter}, both variants settled into a stable following state. The baseline sensor range was adequate, mainly because the relative speed was comparatively lower and the VRU remained in lane. External communication still improves ride comfort by enabling deceleration earlier and resulting in a reduced peak longitudinal jerk.

\begin{table}[t]
\caption{Sustainability Score in Qualitative Analysis}
\setlength{\tabcolsep}{0.8pt}
\renewcommand{\arraystretch}{1.0}
{%
\begin{tabular}{|c|c|c|c|c|c|}
\hline
Scenario & Controller & V2X & Sfty. Score & Eff. Score & Cmft. Score \\ \hline
\multirow{2}{*}{Pedestrian crossing}  & \multirow{2}{*}{CAP} & Disabled & 0.00   & 0.00   & 0.00\\ 
&  & Enabled  & \textbf{1.00}  & \textbf{0.27}  & \textbf{0.59}\\ \hline
\multirow{2}{*}{E-scooter following}   & \multirow{2}{*}{CAP} & Disabled & 1.00  & \textbf{0.67}  & \textbf{0.78}\\  &  & Enabled  & 1.00  & 0.65  & 0.76     \\ \hline
\multirow{2}{*}{Motorcyclist crossing} & \multirow{2}{*}{CAP} & Disabled & 0.00 & 0.00  & 0.00\\ & & Enabled  & \textbf{0.81} & \textbf{0.93} & \textbf{0.89}\\ \hline
\end{tabular}%
}
\label{tab:sustain_score_qual_sim}
\end{table}

Table~\ref{tab:sustain_score_qual_sim} presents the numeric scores for all three scenarios where the vehicle under the CAP without the V2X collides with the crossing pedestrian and motorcyclist, and gets zero due to the collision penalty. In the e-scooter following case, both control methods can decelerate and follow the leading VRU, while CAP with V2X has a marginally lower efficiency and comfort scores due to the more conservative planner design.
These observations also align with the previous findings from \cite{he2024risk} that informative infrastructure, such as V2X communication functionality, can reduce the crash probability in risky VRU interaction scenarios.

\vspace{-1mm}
\section{Conclusion and Future Work}
\vspace{-1mm}
In this paper, we presented a simulation framework that digitizes standardized VRU safety tests, allowing repeatable evaluation of vehicle-VRU interactions. Within the developed environment, we compared a baseline rule-based FSM planner with the same planner augmented by V2X. Our preliminary results confirmed that the early informative assistance can mitigate the risk in various dangerous scenarios.

For future work, we will introduce realistic V2X communication effects, including latency, packet loss, and discontinuities, to assess controller robustness under imperfect connectivity. We also plan to integrate an advanced predictive vehicle motion planner for more resilient collision avoidance capabilities in complex VRU scenarios.

\bibliography{ref}

\begin{thebibliography}{10}

\bibitem{yin2023reliable}
J.~Yin, L.~Li, Z.~P. Mourelatos, Y.~Liu, D.~Gorsich, A.~Singh, S.~Tau, and Z.~Hu, ``Reliable global path planning of off-road autonomous ground vehicles under uncertain terrain conditions,'' {\em IEEE Transactions on Intelligent Vehicles}, vol.~9, no.~1, pp.~1161--1174, 2023.

\bibitem{liu2023metamining}
K.~Liu, L.~Chen, L.~Li, H.~Ren, and F.-Y. Wang, ``Metamining: Mining in the metaverse,'' {\em IEEE Transactions on Systems, Man, and Cybernetics: Systems}, vol.~53, no.~6, pp.~3858--3867, 2023.

\bibitem{li2002automated}
L.~Li and F.-Y. Wang, ``The automated lane-changing model of intelligent vehicle highway systems,'' in {\em Proceedings. The IEEE 5th International Conference on Intelligent Transportation Systems}, pp.~216--218, IEEE, 2002.

\bibitem{nazat2024xai}
S.~Nazat, L.~Li, and M.~Abdallah, ``Xai-ads: An explainable artificial intelligence framework for enhancing anomaly detection in autonomous driving systems,'' {\em Ieee Access}, vol.~12, pp.~48583--48607, 2024.

\bibitem{li2024sora}
X.~Li, Q.~Miao, L.~Li, Q.~Ni, L.~Fan, Y.~Wang, Y.~Tian, and F.-Y. Wang, ``Sora for scenarios engineering of intelligent vehicles: V\&v, c\&c, and beyonds,'' {\em IEEE Transactions on Intelligent Vehicles}, vol.~9, no.~2, pp.~3117--3122, 2024.

\bibitem{wang2019vulnerable}
S.~Wang, E.~Susumu, E.~D. Brown, K.~Cunningham, R.~Kaufman, S.~Horbal, J.~Joyner, L.~Drees, J.~Gainey, B.~Mueller, {\em et~al.}, ``Vulnerable road user injury prevention alliance (vipa): Early data and insights,'' 2019.

\bibitem{CPSC}
{U.S. Consumer Product Safety Commission}, ``{E-Scooter and E-Bike Injuries Soar: 2022 Injuries Increased Nearly 21\%}.'' \url{https://www.cpsc.gov/Newsroom/News-Releases/2024/E-Scooter-and-E-Bike-Injuries-Soar-2022-Injuries-Increased-Nearly-21}, 2023.
\newblock [Online].

\bibitem{euroncap}
EURONCAP, ``{AEB/LSS VRU Systems Test protocol v4.5.1}.'' \url{https://www.euroncap.com/media/80156/euro-ncap-aeb-lss-vru-test-protocol-v451.pdf}, 2024.
\newblock [Online].

\bibitem{xu2022safebench}
C.~Xu, W.~Ding, W.~Lyu, Z.~Liu, S.~Wang, Y.~He, H.~Hu, D.~Zhao, and B.~Li, ``Safebench: A benchmarking platform for safety evaluation of autonomous vehicles,'' {\em Advances in Neural Information Processing Systems}, vol.~35, pp.~25667--25682, 2022.

\bibitem{karpinski2023comparison}
E.~Karpinski, E.~Bayles, L.~Daigle, and D.~Mantine, ``Comparison of motor-vehicle involved e-scooter fatalities with other traffic fatalities,'' {\em Journal of safety research}, vol.~84, pp.~61--73, 2023.

\bibitem{saleh2017vru}
K.~Saleh, M.~Hossny, and S.~Nahavandi, ``Towards trusted autonomous vehicles from vulnerable road users perspective,'' in {\em 2017 Annual IEEE International Systems Conference (SysCon)}, pp.~1--7, 2017.

\bibitem{chenVRU}
H.~Chen, Y.~Liu, C.~Hu, and X.~Zhang, ``Vulnerable road user trajectory prediction for autonomous driving using a data-driven integrated approach,'' {\em IEEE Transactions on Intelligent Transportation Systems}, vol.~24, no.~7, pp.~7306--7317, 2023.

\bibitem{son2019simulation}
T.~Duy~Son, A.~Bhave, and H.~Van~der Auweraer, ``Simulation-based testing framework for autonomous driving development,'' in {\em 2019 IEEE International Conference on Mechatronics (ICM)}, vol.~1, pp.~576--583, 2019.

\bibitem{marko2019scenario}
N.~Marko, J.~Ruebsam, A.~Biehn, and H.~Schneider, ``Scenario-based testing of adas-integration of the open simulation interface into co-simulation for function validation.,'' in {\em SIMULTECH}, pp.~255--262, 2019.

\bibitem{solmaz2019novel}
S.~Solmaz and F.~Holzinger, ``A novel testbench for development, calibration and functional testing of adas/ad functions,'' in {\em 2019 IEEE International Conference on Connected Vehicles and Expo (ICCVE)}, pp.~1--8, IEEE, 2019.

\bibitem{shao2019evaluating}
Y.~Shao, M.~A.~M. Zulkefli, Z.~Sun, and P.~Huang, ``Evaluating connected and autonomous vehicles using a hardware-in-the-loop testbed and a living lab,'' {\em Transportation Research Part C: Emerging Technologies}, vol.~102, pp.~121--135, 2019.

\bibitem{he2023simulation}
Z.~He and L.~Li, ``Simulation framework for vehicle and electric scooter interaction,'' in {\em 2023 IEEE 26th International Conference on Intelligent Transportation Systems (ITSC)}, pp.~4479--4484, 2023.

\bibitem{he2024sustain}
Z.~He, Y.~Guo, and L.~Li, ``Sustainability evaluation in vehicle and electric scooter interaction,'' in {\em 2024 IEEE 27th International Conference on Intelligent Transportation Systems (ITSC)}, pp.~2786--2792, 2024.

\bibitem{2secrule}
{New York State Department of Motor Vehicles}, ``Driver's manual - chapter 8: Defensive driving.'' \url{https://dmv.ny.gov/about-dmv/chapter-8-defensive-driving}, 2011.
\newblock [Online].

\bibitem{he2023vpi}
Z.~He, L.~Capito, K.~Redmill, F.~{\"O}zg{\"u}ner, and {\"U}.~{\"O}zg{\"u}ner, ``Risk analysis for vehicle--pedestrian interaction with extended sensing,'' {\em Towards Human-Vehicle Harmonization}, vol.~3, p.~65, 2023.

\bibitem{lobo2024adaptive}
S.~Lobo, A.~Festag, and C.~Facchi, ``Adaptive message prioritization: How to prioritize vru awareness messages in a congested v2x network,'' in {\em 2024 IEEE 27th International Conference on Intelligent Transportation Systems (ITSC)}, pp.~423--430, 2024.

\bibitem{he2024rampcast}
Z.~He, A.~Mathew, A.~Ingale, J.~Zhou, L.~Li, F.~Li, Y.~Chen, J.~Sturdevant, and E.~Cox, ``Ramp cast: Traffic management geocast study in indiana highways using c-v2x technology,'' in {\em 2024 IEEE 27th International Conference on Intelligent Transportation Systems (ITSC)}, pp.~2104--2109, 2024.

\bibitem{he2024risk}
Z.~He, Y.~Guo, Y.~Chen, B.~King, and L.~Li, ``Risk analysis in vehicle and electric scooter interaction,'' in {\em 2024 IEEE Intelligent Vehicles Symposium (IV)}, pp.~1316--1322, 2024.

\end{thebibliography}

\end{document}